\documentclass{emulateapj}

\begin{document}

\title{The Kilodegree Extremely Little Telescope (KELT): A Small Robotic
Telescope for Large-Area Synoptic Surveys}
\author 
{Joshua Pepper\altaffilmark{1}, Richard W. Pogge\altaffilmark{1}, 
D.~L. DePoy\altaffilmark{1}, J.~L. Marshall\altaffilmark{1}, K.~Z. Stanek\altaffilmark{1},
Amelia M. Stutz\altaffilmark{2}, Shawn Poindexter\altaffilmark{1}, Robert 
Siverd\altaffilmark{1}, Thomas P. O'Brien\altaffilmark{1}, 
Mark Trueblood\altaffilmark{3}, \& Patricia Trueblood\altaffilmark{3}}
\altaffiltext{1}{The Ohio State University Department of Astronomy, 
4055 McPherson Lab, 140 West 18th Avenue, Columbus, OH 43210-1173}
\altaffiltext{2}{Department of Astronomy, University of Arizona, 933 N. Cherry Avenue, Tucson, AZ 85721-0065}
\altaffiltext{3}{Winer Observatory, P.\ O.\ Box 797, Sonoita, AZ 85637-0797 }
\email{pepper@astronomy.ohio-state.edu}

\begin{abstract}
The Kilodegree Extremely Little Telescope (KELT) project is a survey for 
planetary transits of bright stars.  It consists of a small-aperture, 
wide-field automated telescope located at Winer Observatory near 
Sonoita, Arizona.  The telescope surveys a set of 26\degr$\times$26\degr\ 
fields, together covering about 25\% of the Northern sky, targeting stars in the 
range of $8<V<10$ mag, searching for transits by close-in Jupiters.  This paper describes the system hardware and software 
and discusses the quality of the observations.  We show that KELT is able to 
achieve the necessary photometric precision to detect planetary transits 
around solar-type main sequence stars.
\end{abstract}
\keywords{Astronomical Instrumentation}

\section{Introduction} \label{sec:intro}

The scientific value of planetary transits of bright stars is well known -- for 
a comprehensive review see \citet{char07}.  These transits 
provide the opportunity to study the internal structure of 
planets \citep{g05}, their atmospheric composition \citep{char02}, spin-orbit 
alignment \citep{gw07}, and the presence of rings or 
moons \citep{bf04}.  Radial-velocity (RV) surveys have searched the 
brightest stars in the sky for planets and are probing increasingly 
fainter stars.  Even with significant multiplexing, however, RV surveys 
are not able to search large numbers of stars fainter than $V\sim8$ mag.  To 
find planets around fainter stars, transit surveys are more suitable, and 
a number of such surveys are underway.  These surveys typically have wide 
fields of view and small apertures to simultaneously monitor tens or 
hundreds of thousands of stars.  These surveys have so far 
discovered six planets transiting bright stars
\citep{alonso04,bakos07,cam07,mc06,od06}.  In
order to discover more such scientifically valuable systems, we have
begun a survey to discover transiting planets.

The Kilodegree Extremely Little Telescope (KELT) is designed to meet the 
objectives described in \citet{pep03} for a wide field, small-aperture 
survey for planetary transits of bright stars.  That paper derives a model 
for the ability of a given transit survey to detect close-in giant planets (i.e. ``Hot Jupiters''), and determines 
an optimal survey configuration for targeting $8<V<10$ mag main-sequence stars.  Based on 
the model of \citet{pep03}, we expect to detect roughly four transiting planets 
with the KELT survey.

The KELT system has two 
different observing configurations.  The
primary configuration is a ``Survey Mode'' designed for wide area
coverage, either in large strips or an all-sky survey, with the
goal of covering broad sections of the sky with a large field of 
view, at a cadence of a few minutes 
on a nightly basis throughout most of the observing year.  This mode
implements the primary scientific driver of KELT and gives this telescope 
a wider field of view and targets a brighter magnitude range than other 
transit surveys of its kind.  The second
configuration is a ``Campaign Mode'' that uses a smaller field of view
and is designed to conduct short duration intensive observing campaigns
on specific fields.  In Campaign Mode we have undertaken a 74-day
campaign towards the Praesepe open cluster.

In this paper, we describe the instrumentation, deployment, and operations of 
KELT (\S \ref{sec:overview}); we characterize the performance of the different
components and the overall system in the field (\S \ref{sec:performance}); we 
show how precise our photometry is (\S \ref{sec:relphot}); and we provide 
examples of lightcurves for variable stars and transit-like 
events (\S \ref{sec:lcurves}). 

\section{KELT System Overview} \label{sec:overview}

\subsection{Instrument} \label{sec:instr}

KELT consists of an optical assembly (CCD detector, medium-format camera
lens, and filter) mounted on a robotic telescope mount.  A dedicated
computer is used to control the telescope, camera, observation
scheduling, and data archiving system tasks.  One goal in assembling
KELT was to use as many off-the-shelf components and software packages
as possible to speed the development.

The KELT detector is an Apogee
Instruments\footnote{http://www.ccd.com} AP16E thermoelectrically cooled
CCD camera.  This camera uses the Kodak KAF-16801E front-side
illuminated CCD with $4096 \times 4096$ $9\mu$m pixels (36.88 $\times$
36.88\,mm detector area) and has peak quantum efficiency of $\sim$65\% at
600nm.  The AP16E uses a PCI card and cable to control the camera and
thermoelectric cooler (TEC).  According to the camera specifications and 
confirmed by laboratory testing, the camera is operated at a conversion
gain of 3.6\,electrons/ADU and delivers a measured readout noise of
$\sim$15\,e$^-$.  The device is read out at 14-bit resolution 
at 1.3\,MHz, which gives a full-frame readout time of $\sim$30\,seconds.  The 
CCD specifications claim a full-well depth of $\sim$100,000\,e$^-$, but 
the 14-bit ADC saturates at 16383 ADU ($\sim$59,000 e$^-$).  The TEC can cool the
device to $\sim$30\degr\,C below the ambient air temperature.  Nominal
dark current is $0.1-0.2$\,e$^{-}$\,pixel$^{-1}$\,sec$^{-1}$ at an
operating temperature of $-10$\degr\,C (typical for $20$\degr\,C ambient
air temperature).

We use two different lenses with KELT.  For the wide-angle survey mode,
we use a Mamiya 645 80\,mm f/1.9 medium-format manual-focus lens with a
42\,mm aperture.  This lens provides a roughly 23\arcsec\,pix$^{-1}$
image scale and a 26\degr$\times$26\degr\ field of view.  To
provide a narrow-angle campaign mode, we use a Mamiya 645 200\,mm f/2.8
APO manual-focus telephoto lens with a 71\,mm aperture.  
This provides a roughly 9\farcs5 pix$^{-1}$ image scale and effective
10\fdg8$\times$10\fdg8 field of view.  Both lenses 
have some vignetting at the corners, and the image quality declines 
toward the outer part of the field, so the effective field of view is 
circular (see \S \ref{sec:datared} and \S \ref{sec:imqual} for details).

\begin{figure}
\plotone{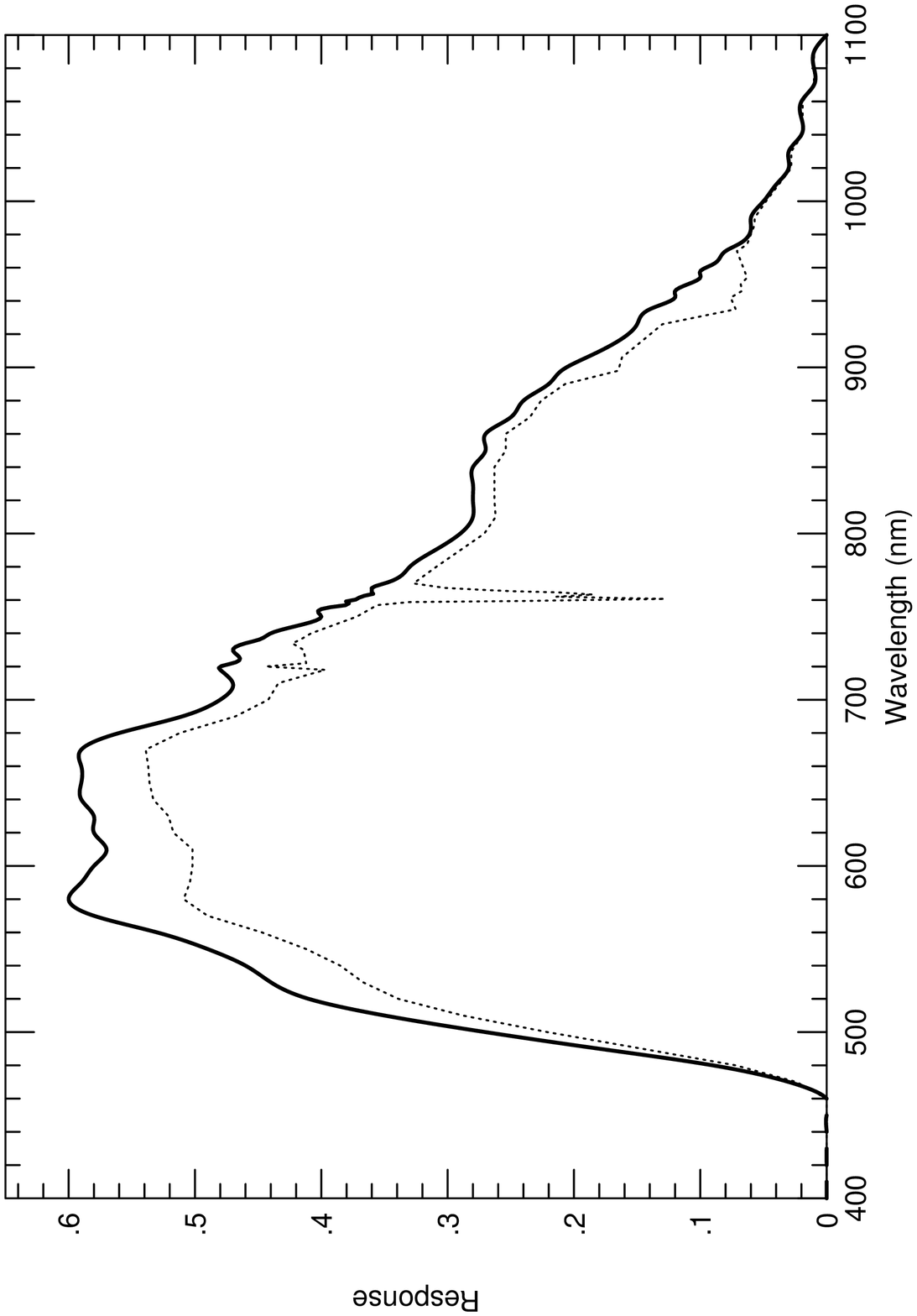}
\caption{Calculated response function of the KELT CCD camera
and Kodak \#8 Wratten filter.  The dashed curve is the response function
including atmospheric transmission at Winer Observatory for 1.2
airmasses.  This response function does not include the transmission of
the camera lenses.}
\label{fig:bpass}
\end{figure}

To reject the mostly-blue background sky without greatly diminishing the
sensitivity to stars (which are mostly redder than the night sky), we
use a Kodak Wratten \#8 red-pass filter with a 50\% transmission point
at $\sim$490\,nm (the filter looks yellow to the eye), mounted in 
front of the KELT lens.  The calculated
response function of the KELT CCD and filter is shown in
Figure\,\,\ref{fig:bpass}.  The transmission function for the Wratten \#8
filter is taken from the Kodak Photographic Filters Handbook
\citep{kodak98}, and the quantum efficiency curve for the Kodak
KAF-16801E CCD was provided by the Eastman Kodak Company.  The
effect of atmospheric transmission on this bandpass is estimated for 
1.2 airmasses at the altitude of Winer Observatory (1515\,m) using the
Palomar monochromatic extinction coefficients \citep{hl75}, which are for an
altitude of 1700\,m.  We did not estimate
in detail the atmospheric water vapor or O$_2$ extinction terms, as
these are not important for our application.  The effective wavelength of
the combined Filter+CCD response function (excluding atmospheric effects)
is 691nm, with an effective width of 318nm, computed following the
definition of \citet{sch83}.  This results in an effective bandpass that is 
equivalent to a very broad R-band filter.

The optical assembly (camera+lens+filter) is mounted on a Paramount ME
Robotic Telescope Mount manufactured by Software
Bisque\footnote{http://www.bisque.com}.  The Paramount is a
research-grade German Equatorial Mount designed specifically for robotic
operation with integrated telescope and camera control.  According to 
manufacturer's ratings the periodic tracking error of the mount before 
correction is $\pm5$\arcsec.  That is smaller than the large pixels of 
the KELT camera and therefore does not affect our observations.  The mount 
can carry an instrument
payload of up to 75\,kg, more than sufficient for the KELT camera, which 
weighs approximately 9\,kg.  The
Paramount is installed on a stock 91\,cm high steel pier using a stock
base adapter plate.  The Paramount provides us with a robust, complete
mounting solution for our telescope.  The optical assembly is mated to
the Paramount using a custom mounting bracket that mounts directly on
the Paramount's Versa-Plate mounting surface.

The CCD camera and mount are controlled by a PC computer located at the
observing site that runs the Windows XP Professional operating system
and the Bisque Observatory Software Suite from Software Bisque.  There
are three main applications that we use for KELT: {\tt TheSky} to
operate the Paramount ME (pointing and tracking); {\tt CCDSoft} CCD
camera control software to operate the AP16E camera; and {\tt Orchestrate}
scripting and automation software to integrate the operation of {\tt
TheSky} and {\tt CCDSoft}.  {\tt Orchestrate} provides a simple
scripting interface that lets us control all aspects of a night's
observing with a single command script.  This software lets us prepare
an entire night's observing schedule and upload it to the KELT control
computer during the afternoon.  A scheduled task on the computer starts up 
the system at sunset and loads the observing script into {\tt
Orchestrate}, after which the system runs unattended for the entire
night, weather permitting.

The observing site provides AC power and Internet connectivity.  To
ensure clean AC power for the KELT telescope and control computer at the
observing site (\S\ref{sec:site}) we use a 1500VA Powerware 9125
Uninterruptable Power Supply (UPS).  This filters the line power and
protects the system against surges or brief power outages.  The control
computer is connected to the Internet through the observing site's
gateway, and its internal clock is synchronized with the network time
servers at the Kitt Peak National Observatory using the {\tt Dimension
4} network time protocol (NTP) client\footnote{v5.0 from Thinking Man Software
(www.thinkman.com/dimension4/)}.
This ensures sufficiently accurate timing for telescope pointing and
time-series photometry.  Since we observe at few-minute cadences, we can
tolerate few-second timing precision, which is easily within the typical 
performance of {\tt Dimension 4} on the available T1 connection.

\subsection{Observatory Site} \label{sec:site}

The KELT telescope has been installed at the Irvin M. Winer Memorial
Mobile Observatory\footnote{http://www.winer.org} near Sonoita, Arizona.
Located at N 31\degr39\arcmin53\arcsec, W 110\degr36\arcmin03\arcsec, approximately 
50 miles southeast of Tucson at an elevation of 4970\,feet
(1515\,meters), Winer Observatory has a dedicated observing
building with a 25$\times$50-foot (7.6$\times$15-meter) roll-off enclosure and provides site
and maintenance services.  Winer currently hosts four robotic 
telescopes, including KELT.  The KELT
telescope pier is bolted to the concrete floor, and its control cables
are run into a nearby telescope control room that houses the control
computer and UPS.  Winer also provides Internet access (currently via a dedicated
DS/T1 line, but early in the project the site used a slower ISDN link), that allows
us to remotely login to the control computer via a secure gateway.

The weather conditions at Winer Observatory are roughly as good as comparable 
sites in southern Arizona; about $60\%$ of all observing time is usable, 
with half of that time being measurably photometric.  Since our PSFs are 
between 2 and 3 pixels, and the pixel scales are 9\farcs5 and 23\arcsec \,(for 
the 200\,mm and 80\,mm lenses, respectively), atmospheric seeing variations 
(which are on the scale of arcseconds) are not a factor in our observations.

\subsection{Observing Operations}\label{sec:operations}

Observations with KELT are carried out each non-clouded night using command
scripts for pre-programmed, robotic operation; we do not undertake any
remote real-time operations.  The nightly observing scripts are created at Ohio State 
University (OSU) using a script-generation program written in Perl, and then uploaded to
Winer Observatory where they are used by {\tt Orchestrate} to direct the
telescope to observe the specified fields for each night.  Each night has a
different script, and we generally upload scripts in 3-4 week batches
during the main survey season, or more frequently during pointed-target
campaigns.

The suite of software programs we use to control the telescope mount and
camera works well, but has several limitations.  Most
importantly, the {\tt Orchestrate} scripting package does not provide
the built-in ability to program control loops or conditional branching, which is
why we use a Perl program to create the {\tt Orchestrate} scripts we
upload to the control computer.

The scripts start the telescope each night based on the
local clock time.  On nights when the weather is judged to be good
enough for observing, the observatory control computer automatically opens the roof of the
observatory at nautical (12\degr) twilight and the telescope begins observations on schedule.
If the weather is not suitable for observing, on-site personnel
abort the command script.  If the weather appears good at first but degrades during
the night, the observatory computer closes the roof and the personnel abort the script.  All
data acquired are archived automatically at the
end of the night.

When the script is loaded into {\tt Orchestrate}, it first waits until
one hour before astronomical (18\degr) twilight.  At that point the telescope takes
five dark images and five bias images, with the exposure times for the
dark frames the same as the exposure times of the observations for the night.
Once these calibration data are taken, the telescope goes back to sleep
until astronomical twilight, at which point it slews to the first target
field and begins the nightly observing sequence.  Unless the weather turns bad, 
prompting the roof to close, observations continue until astronomical dawn.  At this
time the telescope is slewed to its stowed position, and five dark
images and five bias images are acquired, ending observing for the
night.  (See \S \ref{sec:datared} for a discussion of KELT flatfields.)

We have so far used KELT in two distinct operating modes: campaign 
mode, using the 200\,mm lens, and survey mode, using the 80\,mm 
lens.  In campaign mode, the telescope intensively observes
a single target field for an entire night.  In survey mode, the telescope
observes a number of fields that are equally spaced around the sky at
2$^h$ intervals of Right Ascension centered 
on Declination +31\degr 39\arcmin\, (the latitude of Winer).

In campaign mode, the observing script instructs the telescope to wait 
until the target field is above 2 airmasses, and to then observe the 
field continuously until it sinks below 2 airmasses or astronomical 
dawn, whichever happens first.  In survey mode, the telescope begins 
observing after astronomical twilight, tiling between two fields at a 
time.  New fields are observed as they rise above 1.4 airmasses.  
Provisions for Moon avoidance are built into the scripts to prevent
 observations of any survey field when it is within 45\degr\ (two field 
widths from field center) of the Moon.

An operational complication arises because the KELT mount is a German
Equatorial design.  This means that fields observed East of the Meridian
are rotated by 180\degr\ relative to fields observed West of the
Meridian.  The practical effect is that we
must separately reduce photometry for fields taken in East and West
orientations, especially when we use difference-imaging photometry.  To
avoid the complication of creating two separate data pipelines to reduce
data taken in survey mode, we instead observe only fields in the Eastern 
part of the sky.  We lose some observation time due to periods when 
the Moon is within 45\degr\ of all fields in the East 
above 1.4 airmasses, leading to downtime when no field is available that
meets the observing criteria.  In those situations, the scripts instruct
the telescope to pause observing until the next field becomes available.
Overall, the loss to Moon avoidance reduces the total amount of data acquired by
$\sim$10\%.

\subsection{Data Handling and Archiving}\label{sec:datahandle}

Data acquired by the camera are immediately written to a hard drive in
the control computer, logged, and then copied to one of two 250\,GB
external hard drives attached to the control computer by a USB 
interface.  At the end of the night, all new images are automatically 
duplicated onto the other external hard drive.  During long
winter nights, the telescope can take as many as 500-600 images per
night, depending on the exposure time, filling the drives every two 
weeks.  Normally, however, bad
weather and downtime due to Moon avoidance reduce the actual observing
rate, so it typically takes 3-4 weeks for both storage drives to reach
capacity.

Data quality is monitored daily using automated scripts running on the
Windows observing computer at Winer.  At the end of the night, a Perl
script selects three images from the beginning, middle, and end of the
night and uploads them to a computer at OSU.  These sample images are
analyzed for basic statistics: mean, median, and modal sky value and the
mean FWHM of stars measured across the images, and then visually
inspected to ensure that the camera and mount are operating correctly.

When the external drives approach full capacity, they are disconnected
from the computer.  One of the drives is a hardened 
drive made by Olixir Technologies (their Mobile DataVault) that serves
as the transport drive.  This drive is shipped via FedEx to the OSU
Astronomy Department in Columbus, Ohio, in a cushioned transport 
box, as bandwidth limitations at Winer Observatory preclude online data transfer. The
second drive is a conventional external USB drive without special
mobile packaging made by Maxtor Corporation which serves as the backup
drive and which is stored at Winer Observatory.  Until the transport
drive arrives at OSU and its data have been successfully copied and
verified, the backup drive at Winer is stored and left idle.  In
the event a transport drive arrives damaged, data from the backup drive
will be copied to another transport drive and shipped.  In the meantime,
the removed drives are replaced with two others of each type and operations 
continue.  At any given time, we have four external disk drives in use:
two drives in operation, one in transit, and one stored as a backup. For
transferring hundreds of gigabytes of images every few weeks
during the prime observing season, this procedure has proven to be very
reliable and efficient.  To date, out of dozens of drive shipments, we 
have lost only two drives to damage
in transit, with no loss of data (both arrived damaged at Winer after
their data were retrieved and copied at OSU).

When a transport drive arrives at OSU, the drive is connected to our main
data storage computer and all of the images are copied and run though a
series of data quality checks.  The image files are renamed, replacing
the cumbersome default file name created by the {\tt CCDSoft}
application with a name indicating the field observed, the UTC date, and
an image number.  The images are then analyzed to measure image quality
(modal sky and mean FWHM).  If the modal sky value is above 800 ADU (due
to moonlight, cloud cover, or other ambient light sources), the image is 
discarded.  The cutoff at 800 ADU was determined using two months of 
representative data showing that images compromised by excessive light 
contamination consistently had sky values above that level and were 
unsuitable for photometry.  The cutoff is high, with many poor images 
well below the cutoff, but we choose to be conservative about eliminating 
images early in the reduction process. The images that pass the initial 
filter on sky values, and the trimmed sections of the bad images, are 
stored on a multi-terabyte RAID storage array at OSU, providing data 
redundancy and fast access for data reduction and analysis.

\subsection{Data Reduction}\label{sec:datared}

Here we describe the data reduction process in brief, for the purposes
of evaluating the performance of the KELT camera.  Detailed
descriptions of the reduction process will be included in an upcoming paper on the
scientific results of KELT observing campaigns (J. Pepper, in preparation).

The data reduction pipeline consists of three steps.  First, we 
process the images by subtracting dark frames and dividing by a flatfield.  Second, we
identify all the stars in the field and determine their instrumental
magnitudes.  Third, we obtain the photometry on all images using
difference image analysis.

Dark images are created for each night by median-combining 10 dark
images -- five from the beginning and end of each
night.   In early testing we determined that we can treat our dark frames 
as combined dark+bias.  We take bias frames separately to monitor their 
stability, but do not incorporate them into the reduction process because 
the bias has been extremely stable.  In 
cases where dark frames were not taken or 
there were problems with the images, we use good dark frames from nights 
bracketing the observations to create a substitute dark frame.  We confirmed 
that using dark images from nearby nights did not significantly affect the statistics 
of the subtracted images -- our dark images are quite stable from night to night.

The KELT system is challenging to accurately flatfield.  For the 200\,mm lens 
there is a combined decrease in flux of $\sim$18\% between the center of the 
image and the edges, and a decrease of up to $\sim$26\% between the center and 
the corners.  For the 80\,mm lens, the decrease is $\sim$23\% from 
the center to the edge, and $\sim$35\% from the center to the corners. Because 
of the large KELT field of view (10\fdg8 and 26\degr\ for the 200\,mm 
and 80\,mm lenses, respectively), twilight flats are not useful for flatfielding 
since the twilight sky is not uniform on those scales.  Dome flats using a
diffuse screen produced reasonable results with high signal-to-noise 
ratio.  The flats are sufficiently repeatable that we do not need to regularly 
take dome flats.  For relative photometry, the dome flats work adequately, and 
we are able to absolutely flatfield our images to $\sim 5\%$ accuracy.

Once images have been dark-subtracted and flatfielded, we can then
create catalogs of images and measure the brightnesses of stars on the
images.  This photometric analysis is done in two basic steps.  The
first is to create a high-quality reference image for a field by
combining a few dozen of the best images and then use the
DAOPHOT software package \citep{stet87} to identify all of the stars in
the field down to a faint magnitude limit and measure their 
approximate instrumental magnitudes. See \S\ref{sec:relcalib} for the 
details on how the instrumental magnitudes are calibrated to standard photometry.

Once a template and DAOPHOT star catalog with baseline instrumental
magnitudes have been created, the second step is to process the images
with the ISIS image subtraction package \citep{al98,alard00}.  Our reduction 
process is similar to that of \citet{hart04}.  The ISIS
package first spatially registers all of the images to align them with
the reference image.  The reference image is convolved with a kernel for
each image and subtracted, creating a difference image.  The flux for
each star identified on the reference image is then measured on each
subtracted image using PSF-fitting photometry.  Image 
subtraction has been shown in limited tests to 
be equal to or better than other photometric methods for the purposes of 
transit searches \citep{bak06}.  In section \S \ref{sec:relphot} below, we 
provide additional information about the reduction process to obtain relative 
photometry.

\section{Instrument Performance}\label{sec:performance}

In this section we quantify the performance of the KELT system by assessing in turn the
telescope mount, the astrometric quality (geometric image quality), the
image quality (position-dependent PSF), and photometric sensitivity.

\subsection{Telescope Performance}\label{sec:mount}

Since the telescope was installed at Winer in October 2004, the hardware
has performed up to specifications.  There have been no significant
problems with the mount or the control software.  The pointing has not
been perfect: our fields are so large
that minor pointing errors do not significantly affect our scientific
results, but during normal operations the typical intra-night drift is
$\sim$25\arcmin\ in Declination and $\sim$9\arcmin\ in Right Ascension.  
We believe the drift is due to a slight non-perpendicularity between the 
orientation of the camera and the axis of the mount.  While the magnitude of 
the drift seems large, it represents a movement of $\sim$65 pixels, less 
than 2\% of the size of the field.  It does not cause stars to move across 
large portions of the detector, and therefore does not lead to significant 
changes in the PSFs of individual stars.  Since our reduction method utilizes 
image subtraction, we do lose the ability to take good photometry at the edges 
of a field.  However, because of PSF distortions and other 
effects (see \S \ref{sec:imqual}), we already have degraded sensitivity in 
those regions.  Therefore the loss of coverage and sensitivity due to drift 
is quite small.  Future alignment of the telescope will attempt to reduce or 
eliminate the drift.

\subsection{Astrometric Performance}\label{sec:astrometry}

Measurements of the positions of stars from the Tycho-2 Catalog
\citep{hog00} are used to determine the conversion between pixel
coordinates and celestial coordinates on the KELT images.  We use the
{\tt Astrometrix}\footnote{\url{http://www.na.astro.it/$\sim$radovich/wifix.htm}}
package to compute polynomial astrometric solutions for our images
following the procedure described by \citet{cg02}.  To avoid stars that
are saturated on the KELT images, we consider Tycho-2 stars with
magnitudes $9.0 \leq V_{Tyc} \leq 10.0$.  From these we select up to
1000 stars per image.  A first attempt to
compute a global astrometric solution for the entire 4096$\times$4096
image produced large residuals for most of the outer parts of the
detector, with discrepancies between the predicted and actual positions of
Tycho-2 stars of many tens of pixels.  Since our primary goal 
is to convert pixel coordinates
into celestial coordinates on a star-by-star basis, a global solution is
not required.  We instead divide the image into 25 subimages on a
5$\times$5 grid and perform a separate astrometric solution for each
subimage.  A third-order polynomial astrometric fit is computed for
each subimage using {\tt Astrometrix}.  The individual subimage fits
give much better results, with offsets between predicted and measured
positions of catalog stars at the subpixel level except at the extreme
corners of the field.  The subframes overlap by a few tens of pixels, and
fits to stars common to adjacent subimages are consistent at the
arcsecond level.  For the 200\,mm lens, the typical RMS residuals are
$\pm$0\farcs8, or $<10\%$ of the average pixel size of
$\sim$9\farcs5.  The 80\,mm lens has slightly worse RMS
residuals, $\pm5$\arcsec or $\sim$20\% of a pixel size of
22\arcsec, but still well within tolerances for our
purposes.

Having a good astrometric fit to the images permits a quantitative
assessment of the geometric performance of the KELT optics; specifically, 
variation in pixel size and shape across the field.  Since this is
a commercial lens with proprietary optical designs, there is no way to 
determine these propoerties a priori.  This analysis will therefore be useful 
for anyone contemplating using ssimilar systems, and has implicatons for 
the potential use of theis setup - for example, such a camera/lens 
combination cannot be used to construct large-scale image mosaics.  
Furthermore, while sky subtraction deals with this effect, flatfielding does not.

\begin{figure}
\plotone{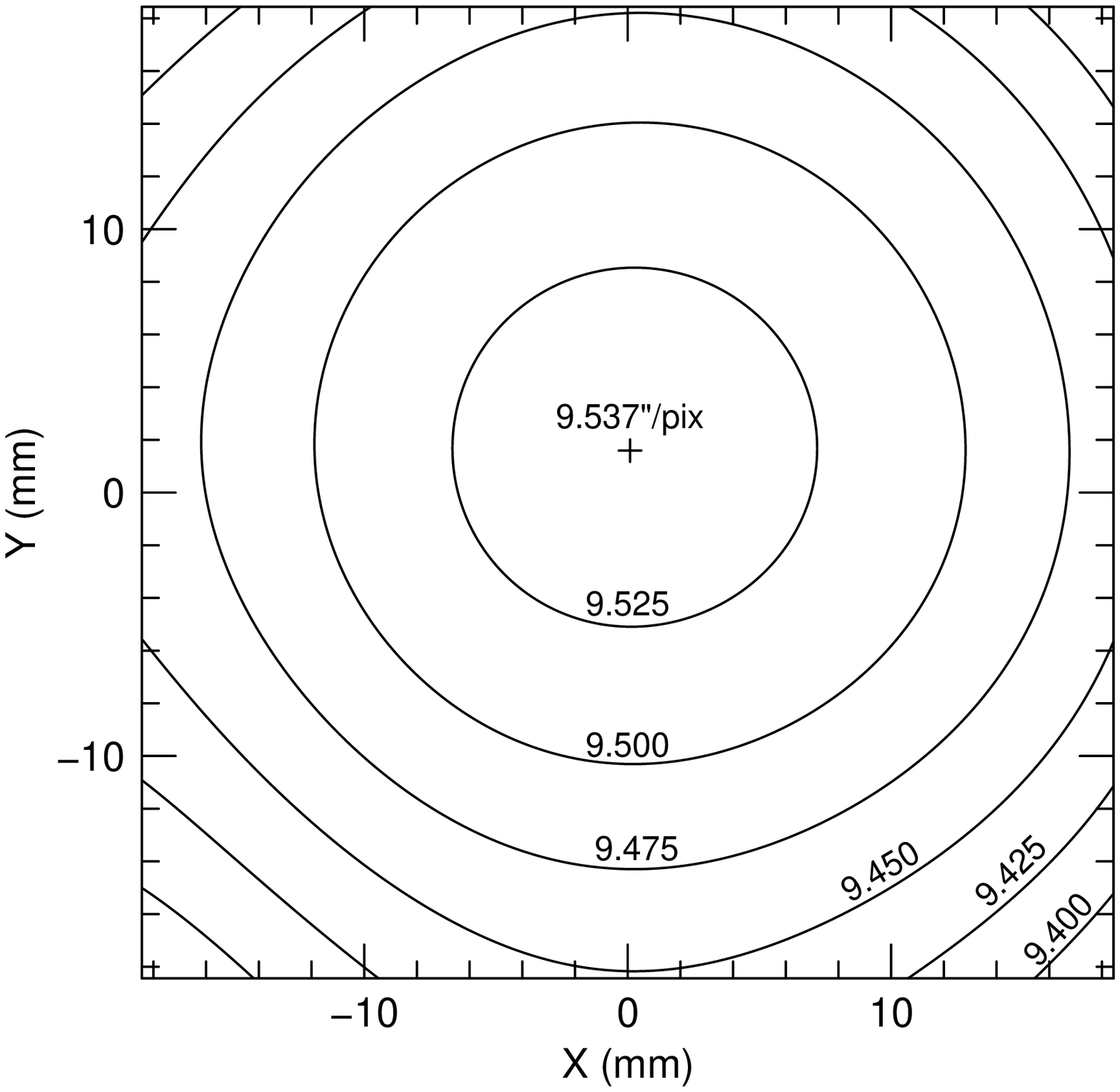}
\caption{Effective pixel scale in arcseconds\,pixel$^{-1}$ for the KELT 200\,mm
camera.  Contours show curves of constant effective pixel scale.  The cross
(+) marks the optical center of the field, where the pixel scale is
9\farcs537\,pix$^{-1}$.}
\label{fig:pscale200}
\end{figure}

\begin{figure}
\plotone{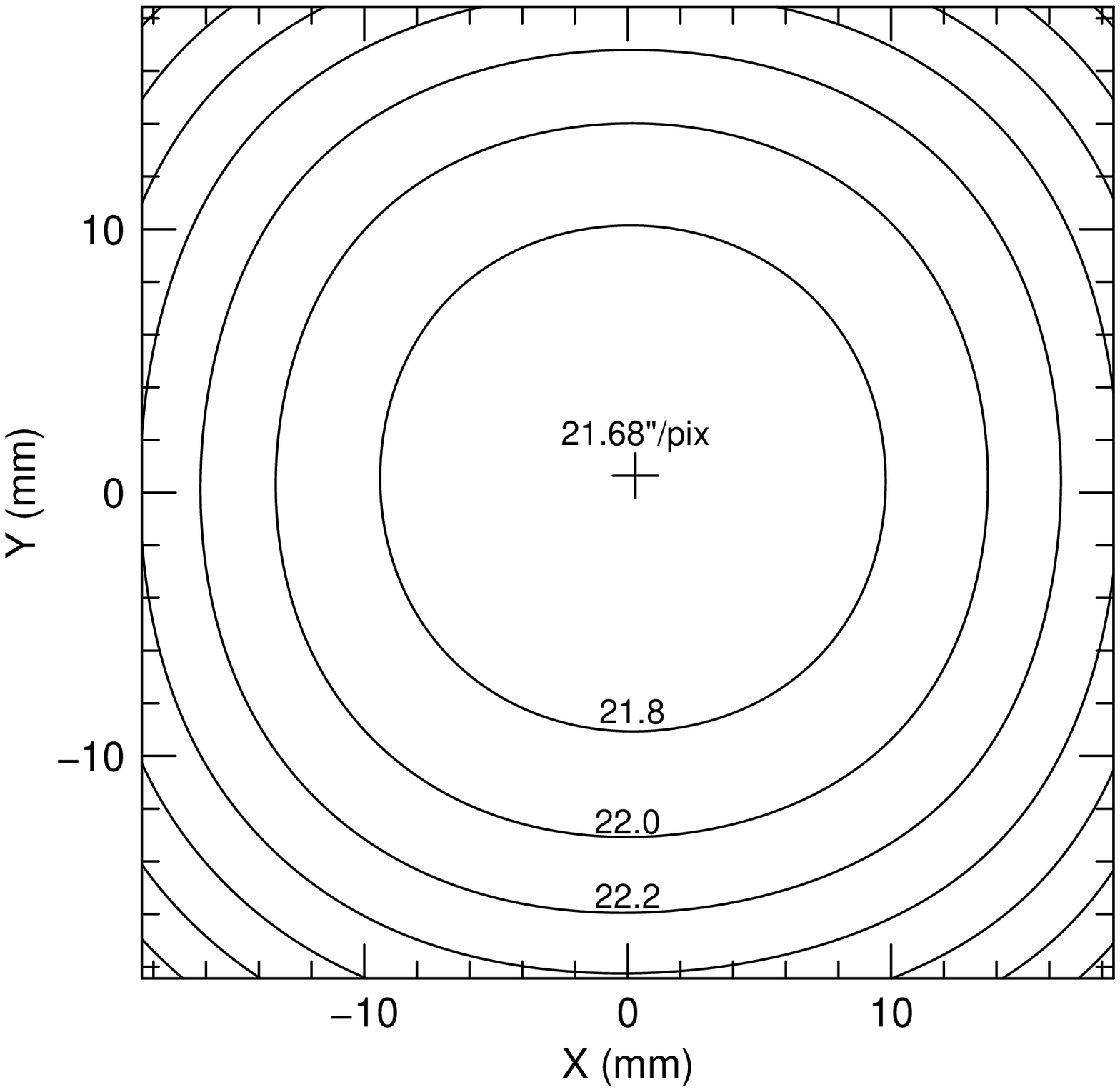}
\caption{Effective pixel scale in arcseconds\,pixel$^{-1}$ for the KELT 80\,mm
camera.  Contours show curves of constant effective pixel scale.  The cross
(+) marks the optical center of the field, where the pixel scale is
21\farcs68\,pix$^{-1}$.}
\label{fig:pscale80}
\end{figure}

For the 200\,mm lens, the effective pixel size decreases by about 1\% from
center-to-edge from 9\farcs537 near the center to
9\farcs450 at the edges of the field
($\sim$9\farcs40 at the corners).  Contours of constant
effective pixel scale are circular and centered on the intersection of
the optical axis of the 200\,mm camera lens and the CCD detector, as
shown in Figure\,\,\ref{fig:pscale200}.  The square CCD pixels do not
perfectly project onto squares on the sky, but slowly distort
systematically away from the center, showing the characteristic
signature of $\sim$0.5\% pincushion distortion, expected for the
manufacturer's typical claims for their telephoto lenses.  As 
Figure\,\,\ref{fig:pscale80} shows, for the
80\,mm lens the effective pixel size decreases by about 3.5\% from
center-to-edge, from 23\farcs19 near the center to
22\farcs40 at the edges of the field (and
$\sim$21\farcs8 at the corners).  This effect is larger than the one 
seen with the 200\,mm lens, consistent with $\sim$2\% pincushion
distortion in this lens, typical of short focal-length wider-angle
lenses.  Contours of constant effective pixel scale are also circular
and centered on the CCD detector.

The effect of the optical distortion is that pixels project onto smaller
effective areas on the sky moving radially outward from the center
of the CCD, making the sky appear non-flat (center-to-edge) at the
$\sim$1.5\% level for the 200\,mm lens and at the $\sim$6\% level for 
the 80\,mm lens.  There are two effects that act together to decrease the background sky
level per pixel as you go radially outward from the center of the image:
the decreasing pixel scale, and hence decreasing pixel
area on the sky, with radius, and increasing vignetting with 
radius.  

\subsection{Image Quality}\label{sec:imqual}

As expected for such an optically fast system, the image PSF varies
systematically as a function of position.  The 
small physical pixel size ($9\mu$m)
implies that the lens optics dominate the PSF, and we are insensitive to changes in
atmospheric seeing.  For both lenses, the systematic patterns in both the image
full-width at half maximum (FWHM) and more refined measures of image
quality (i.e. the 80\% encircled energy diameter $D_{80}$) can be used
to quantitatively assess the position-dependent image quality.

\begin{figure}
\plotone{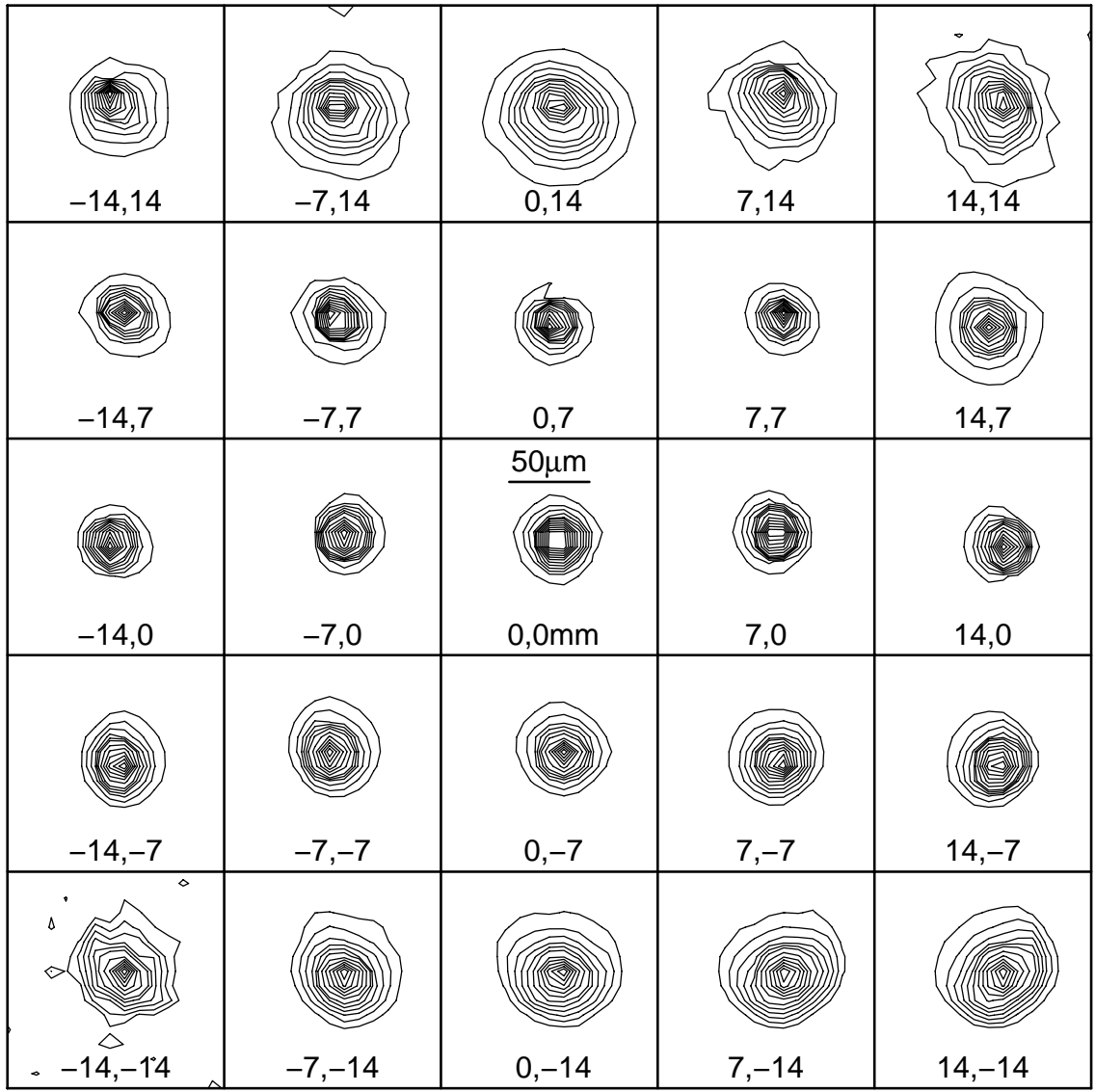}
\caption{Representative stellar PSFs from the 200\,mm telephoto
lens, shown as intensity contours of bright, unsaturated stars taken
with a 5$\times$5 grid pattern on the CCD; the position of the center of each box, relative 
to the center of the image, is indicated in mm.  
Each box is 15\,pixels (135mm)
on a side.  The scale-bar in the center panel indicates 50\,$\mu$m on
the detector (5.5\,pixels).  Contours show levels of
(5,10,15,20,25,30,40,50,60,70,80,90,100)\% of the peak intensity.}
\label{fig:200mmpsf}
\end{figure}

\begin{figure}
\plotone{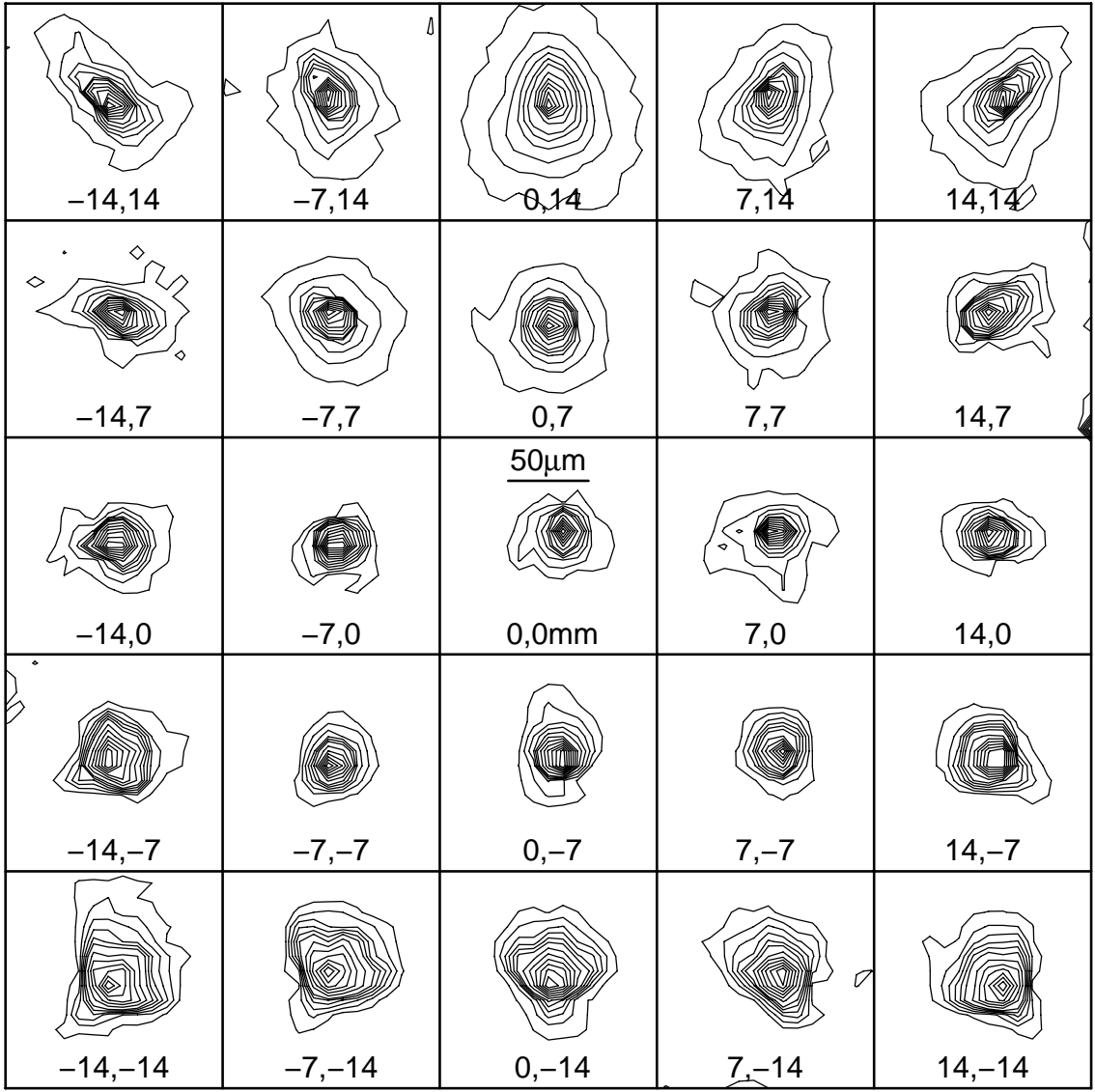}
\caption{Representative stellar PSFs from the 80\,mm wide-angle 
lens, displayed in the same format as in Figure \ref{fig:200mmpsf}.  The images
are more severely distorted at the extreme edges of the field than with the 200\,mm lens.}
\label{fig:80mmpsf}
\end{figure}

For the 200\,mm lens, typical image PSFs have FWHM of
$\sim$1.8--2.9\,pixels, and 80\% encircled energy diameters of
$D_{80}$\,=\,4.7-9\,pixels.  The 80\,mm lens has a similar range of FWHM for
stellar image PSFs, and the 80\% encircled energy diameters range from
$D_{80}$\,=\,6-10\,pixels.  There are significant changes in the
detailed PSF shape across each image from the center to the extreme edges of
the detector.  Figure\,\,\ref{fig:200mmpsf} shows representative stellar
PSFs for a 5$\times$5 grid across the CCD for the 200\,mm lens.  The 80\,mm lens shows more
pronounced distortions, as shown in Figure\,\,\ref{fig:80mmpsf}.  Therefore, 
the 80\,mm lens has a roughly $24^{\circ}$ diameter effective field of view
with reasonably good images and little vignetting, whereas the 200\,mm
lens works well over most of the CCD detector except at the extreme
corners.

\begin{figure}
\plotone{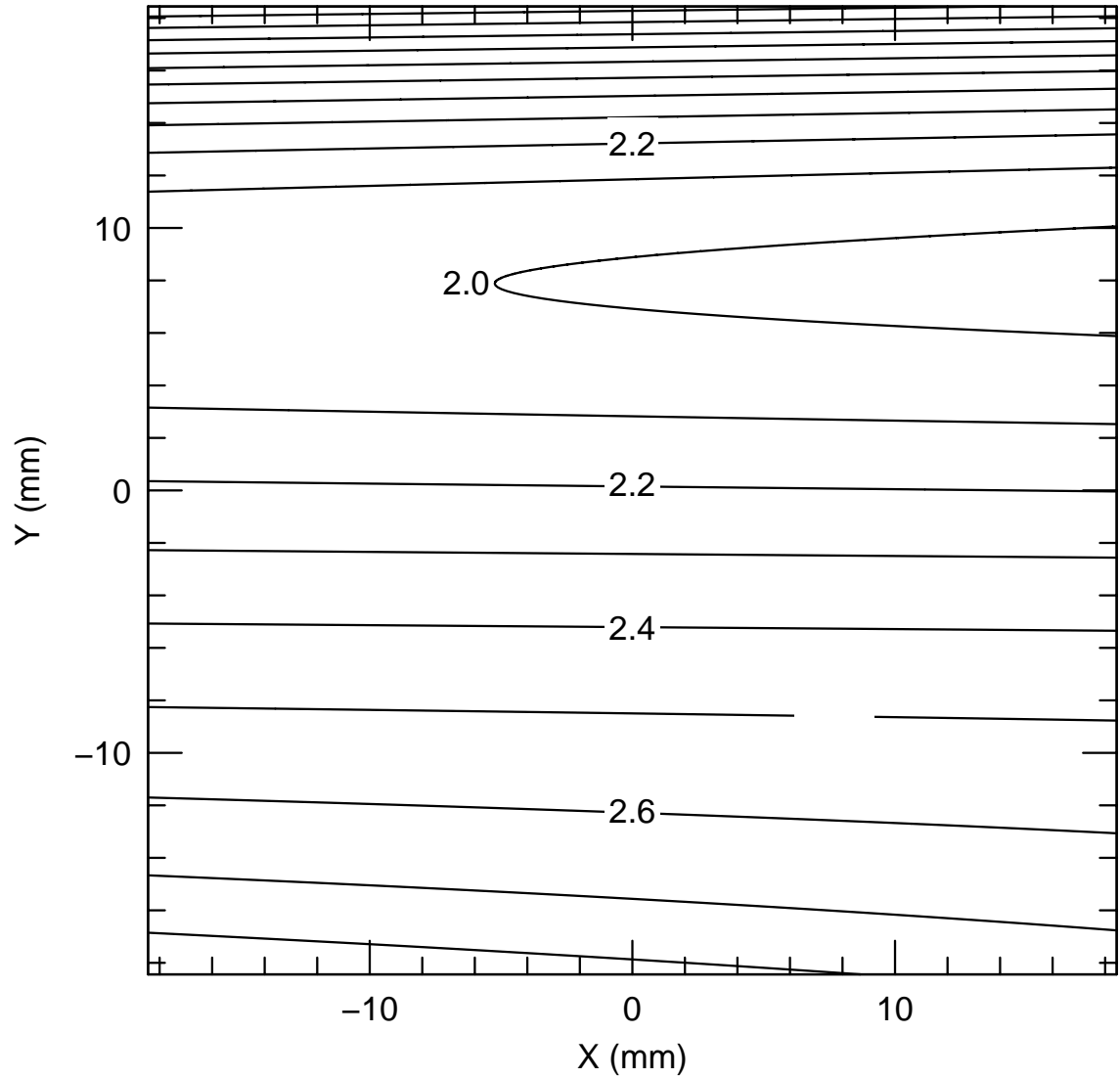}
\caption{Contours of constant FWHM for stellar images in a 
representative 200\,mm lens KELT image, with FWHM given in pixels.  Contours are based
on a smooth polynomial surface fit to measurements of $\sim$1200 bright,
unsaturated stars distributed across the image.  Contour spacing is every 0.1 pixels, 
with particular contour level
values in pixels as indicated.}
\label{fig:fwhm200}
\end{figure}

\begin{figure}
\plotone{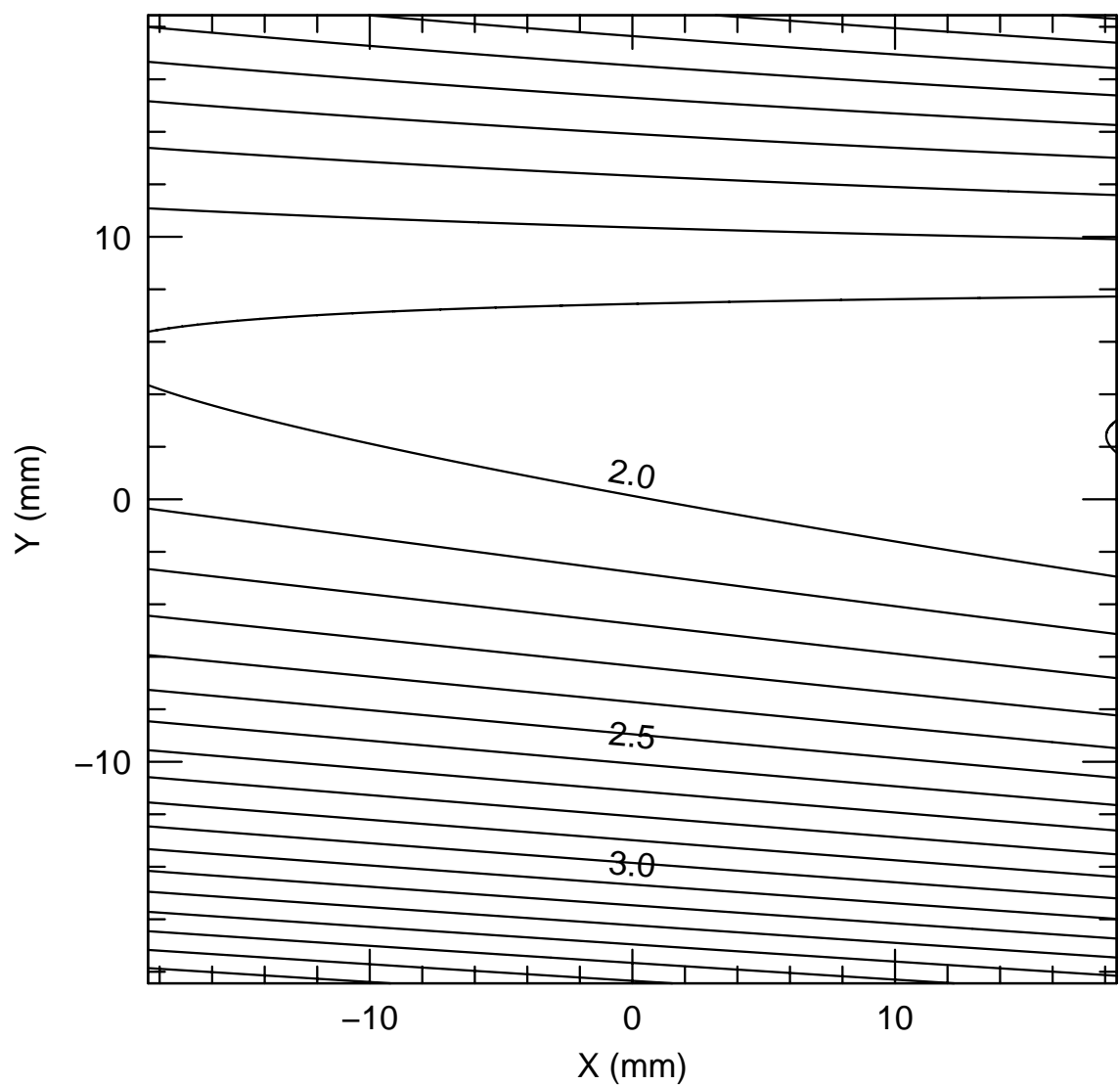}
\caption{Contours of constant FWHM for stellar images in a 
representative 80\,mm lens KELT image, with FWHM given in pixels.  Format is 
the same as in Figure \ref{fig:fwhm200}.  Contours are based on measurements 
of $\sim$2100 bright, unsaturated stars.}
\label{fig:fwhm80}
\end{figure}

\begin{figure}
\plotone{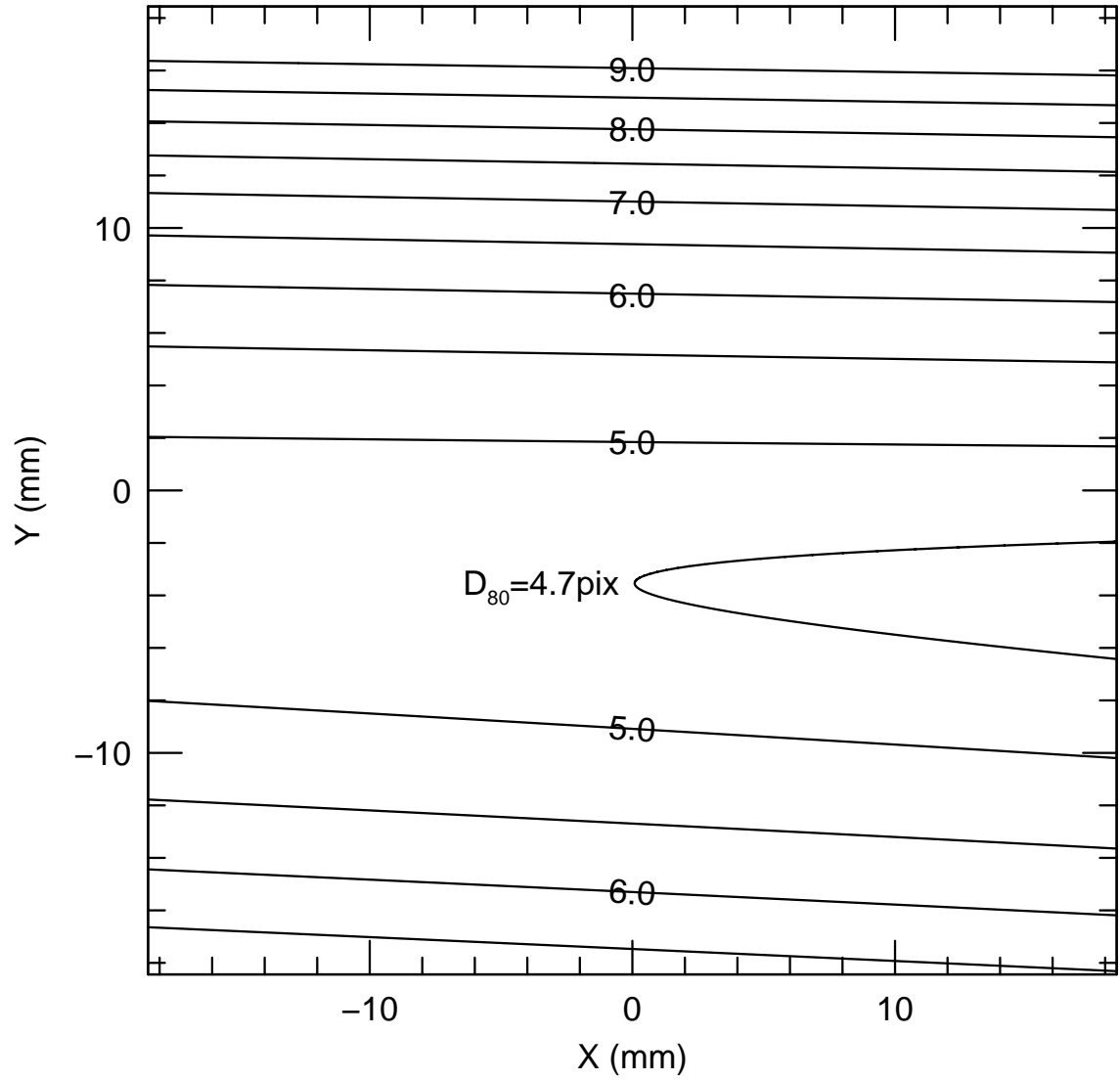}
\caption{Contours of constant 80\% Encircled-Energy Diameter ($D_{80}$) in
pixels for stellar images in a representative 200\,mm lens KELT image.
Contours are based on a smooth polynomial surface fit to measurements of
$\sim$1200 bright, unsaturated stars distributed across the image.}
\label{fig:ee200}
\end{figure}

\begin{figure}
\plotone{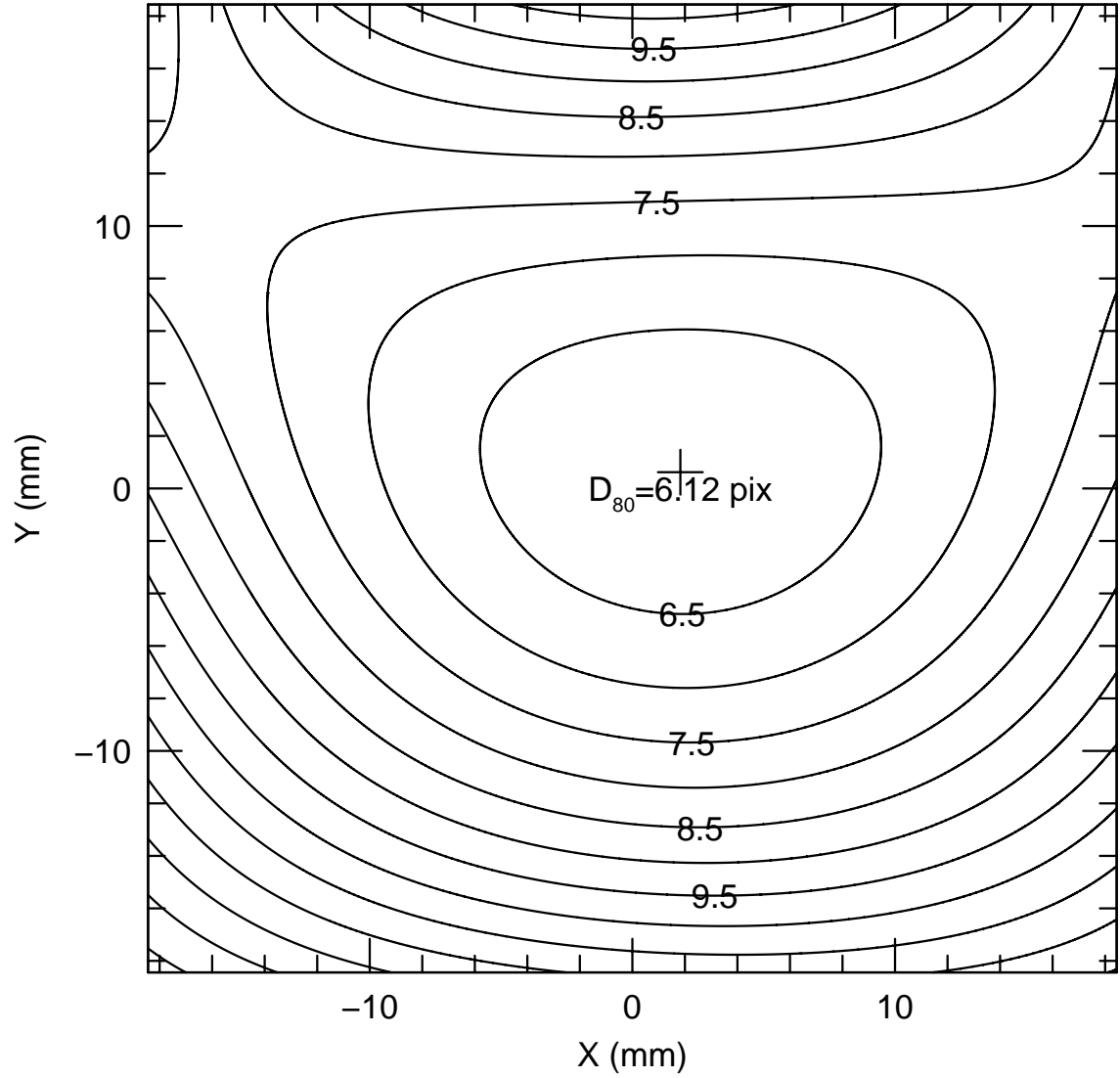}
\caption{Contours of constant 80\% Encircled-Energy Diameter ($D_{80}$) in
pixels for stellar images in a representative 80\,mm lens KELT image.
Contours are based on measurements of
$\sim$2100 bright, unsaturated stars.}
\label{fig:ee80}
\end{figure}

Figures\,\,\ref{fig:fwhm200} and\,\,\ref{fig:fwhm80} show maps of the
image FWHM as a function of position for the 200\,mm and 80\,mm lenses,
respectively, derived from measurements of unsaturated, bright field
stars in representative images.  The most obvious
feature in both is the strong vertical trend in increasing FWHM, with
nearly no differences horizontally.  Because this is seen with both
lenses, which are of very different design, we believe this is because
the CCD is tilted relative to the optical axis.  Because the lens
designs are proprietary, we do not know precisely how much the detector
is tilted, nor the origin of the tilt at present.  This effect could be due to how the
detector is mounted inside the camera, or to the camera/lens mounting
plate.  This apparent field tilt also affects the maps of the 80\%
encircled-energy diameter $D_{80}$, shown in Figures\,\,\ref{fig:ee200} and
\,\ref{fig:ee80}.

The 80\,mm lens has very stable imaging performance over time.  The FWHM
and $D_{80}$ maps derived for images of the same field over a 11-month
period show no significant changes.  We have, however, 
periodically adjusted the focus when working with the telescope, which 
my cause some discontinuities in the data for the survey images.  We 
will explore such effects in upcoming papers.

Unfortunately, the PSF is not stable across the image over time for 
the 200\,mm lens.  While intranight variations in the
FWHM maps are quite small, there are significant changes from
night to night that have no apparent correlation with hour angle,
CCD temperature, or any other physical or environmental parameter for which
we have measurements.  The effect of the changes we see is for the
region of best FWHM (the base of the trough seen in the FWHM map in
Figure\,\,\ref{fig:fwhm200}) to move vertically on the CCD by many
hundreds of pixels.  We do not yet know the cause.  The main effect
is to complicate the difference-imaging reductions of the data.  We will
discuss these complications and their mitigation in the subsequent paper
describing our results for the 200\,mm lens Praesepe cluster observing
campaign.

\subsection{Photometric Sensitivity}\label{sec:abcalib}

Given the nature of the KELT bandpass (see Figure\,\,\ref{fig:bpass}), we
calibrate our instrumental magnitudes to the $R$ band.  We do so by rescaling 
our instrumental magnitudes by a constant, such that 
\begin{equation} \label{equ:pcalib1}
R_K \equiv -2.5\,{\rm log(ADU/sec)} + R_{K,0}\,
\end{equation}
where the instrumental ADU/sec 
is measured using aperture photometry with IRAF, $R_K$ is defined as an approximate 
KELT $R$ magnitude, and $R_{K,0}$ is the zero-point.  We find that the $R_K$ magnitudes 
are within a few tenths of a magnitude of 
standard $R$ band photometry, with the uncertainty dominated by the color 
term.  Since we do not have $V-I$ colors for all our stars, we quote 
observed magnitudes in $R_K$, which can be considered equivalent to Johnson $R$, 
modulo a color term defined by
\begin{equation} \label{equ:pcalib2}
V = R_K + C_{VI}(V-I)
\end{equation}
where $C_{VI}$ is the $(V-I)$ color coefficient, and $V/I$ are in the 
Johnson/Cousins system.  Since we do not have previously measured $R$ magnitudes 
of large numbers of stars in our fields in our magnitude range, we relate $R_K$ to 
known magnitudes by matching stars from our observations to the Hipparcos 
catalog, selecting only stars with measured $V$ and $I$ colors in Hipparcos.  We take 
the mean instrumental magnitude from a set of high-quality images, and match 
the known magnitudes to the mean instrumental magnitudes, using 
Equations \ref{equ:pcalib1} and \ref{equ:pcalib2}.

For the cluster observations with the 200\,mm lens, we select 22 calibration 
stars, and measure their instrumental magnitudes on 76 high-quality images, 
resulting in a magnitude zero-point of $R_{K,0} = 16.38\pm0.06$ mag, 
and $C_{VI}=0.55\pm0.2$.  For the survey observations with the 80\,mm lens, we use 59 
stars on 77 images, resulting in a magnitude zero-point 
of $R_{K,0} = 15.15\pm0.07$ mag, and $C_{VI}=0.5\pm0.3$.  Since the Hipparcos 
stars we use to calibrate our data have $(V-I)$ colors mostly between 0 and 1, we 
expect our calibrations to be less accurate for redder stars.  These measured 
zero-point uncertainties suggest that the flatfield corrections are good to 
within $\sim5\%$ in absolute accuracy.

Tying together the full calibration process, we find that a fiducial $R = 10$ mag
star at the field center with $(V-I) = 0$ has a 
flux of 356 counts per second with the 200\,mm lens, 
and 115 counts per second with the 80\,mm lens.  Scaling those numbers by the 
different aperture sizes of the lenses (71\,mm aperture for the 200\,mm lens and 
42\,mm aperture for the 80\,mm lens), we find that the 200\,mm lens is about 8\% more 
efficient than the 80\,mm lens.

\section{Relative Photometry}\label{sec:relphot}

KELT was designed primarily for precision time-series relative 
photometry (see \citet{eh01} for background).  The crucial test for our instrument 
is the ability to obtain long-term lightcurves with low noise and minimal 
systematics.  A simple test of KELT's photometric performance is to 
examine the root-mean-squared (RMS) of the magnitudes of an ensemble of 
lightcurves as a function of magnitude.  We apply the ISIS image 
subtraction package to samples of our data to obtain relative 
photometry, and measure the statistics of the resulting lightcurves.  Our 
criteria for the ability to detect planetary transits is the presence 
of substantial numbers of stars for which the RMS of the lightcurves are 
below the 2\% and 1\% levels.

\subsection{Difference Imaging Performance}\label{sec:relcalib}

The instrumental magnitudes for the KELT lightcurves are produced through 
a combination of ISIS and DAOPHOT photometry.  This process involves 
some careful conversion between DAOPHOT and ISIS flux measurements -- 
see Appendix B of \citet{hart04} for the details of the conversion.  First, 
we create a reference image by combining a number of high-quality images.  
DAOPHOT measures the instrumental magnitude of the stars on the reference 
image $m_i({\rm ref})$ based on PSF fitting photometry, with the magnitude 
of each star $i$ calculated from the flux 
by $m_i({\rm ref}) \equiv 25 - 2.5\log[f_i({\rm ref})] + C_{ap}$, where $f_i$
is the flux measured by DAOPHOT and $C_{ap}$ is an aperture correction
to ensure that $m_i({\rm ref}) = 25 - 2.5\log(c_i)$, where $c_i$ is the
counts per second from the star in ADU.  ISIS then creates an ensemble of 
subtracted images for the whole data set using the reference.  To derive 
the full light curve, ISIS fits a PSF for each star on each
subtracted image $j$, to obtain a flux $f_i(j)$.  The DAOPHOT-reported
instrumental magnitudes for the reference images serve as the magnitude
baseline for the conversion of ISIS fluxes to magnitudes, where the
magnitude of the $i$th star on the $j$th image is $m_i(j) = m_i({\rm
ref}) - 2.5\log[1 - f_i(j)/f_i({\rm ref})]$.

\begin{figure}
\plotone{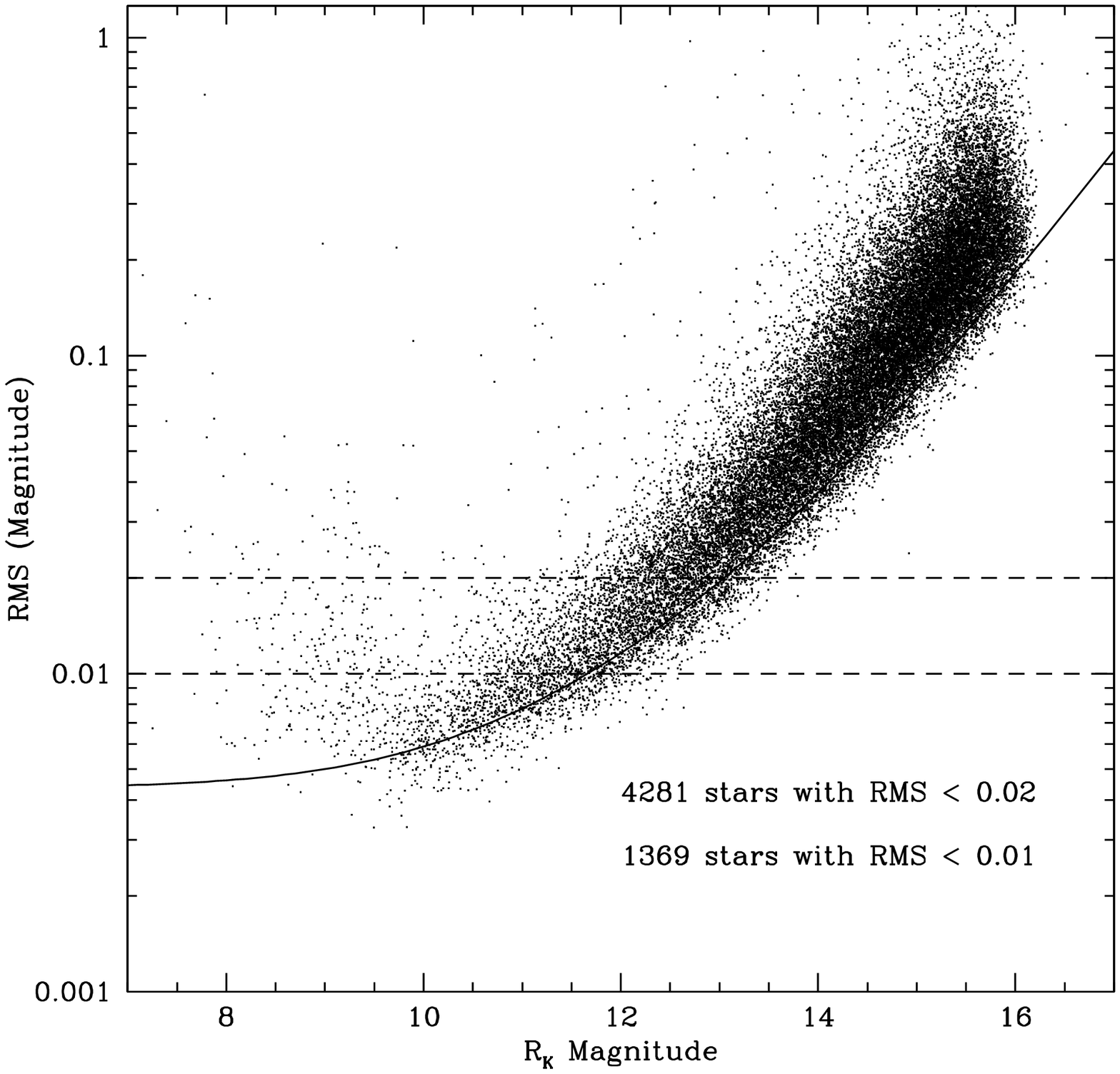}
\caption{RMS plot for one night of data from the 200\,mm lens.  Data are 
shown for 67,674 stars.  The dashed lines show the limits for 1\% and 2\% RMS.  The 
solid line represents a noise model including photon noise and sky noise, along with 
an RMS floor of 0.004 magnitudes.}
\label{fig:crms}
\end{figure}

We then calculate the RMS variation of all the detected stars in both 
the Praesepe data set and for a sample of the survey data.  Because of 
the night-to-night variations in the position of the best image quality 
on the detector with the 200\,mm lens described at the 
end of \S \ref{sec:imqual}, we calculate the RMS for the Praesepe data 
from a single night of observations, to better demonstrate the intrinsic 
instrumental performance.  In Figure\,\,\ref{fig:crms} we plot the 
distribution of RMS versus $R_K$ magnitude for 67,674 stars on 32 images 
with 60-second exposures from one night.  With this lens and exposure 
time, we obtain photometry of stars in the magnitude 
range $R_K = 8-16$ mag.  For stars brighter than about $R_K = 9.5$\,mag, 
systematics begin to dominate the light curves, mostly due to 
saturation.  Out of the 67,674 stars, 
4,281 have RMS\,$<0.02$\,mag, and 1,369 have RMS\,$<0.01$\,mag.  

\begin{figure}
\plotone{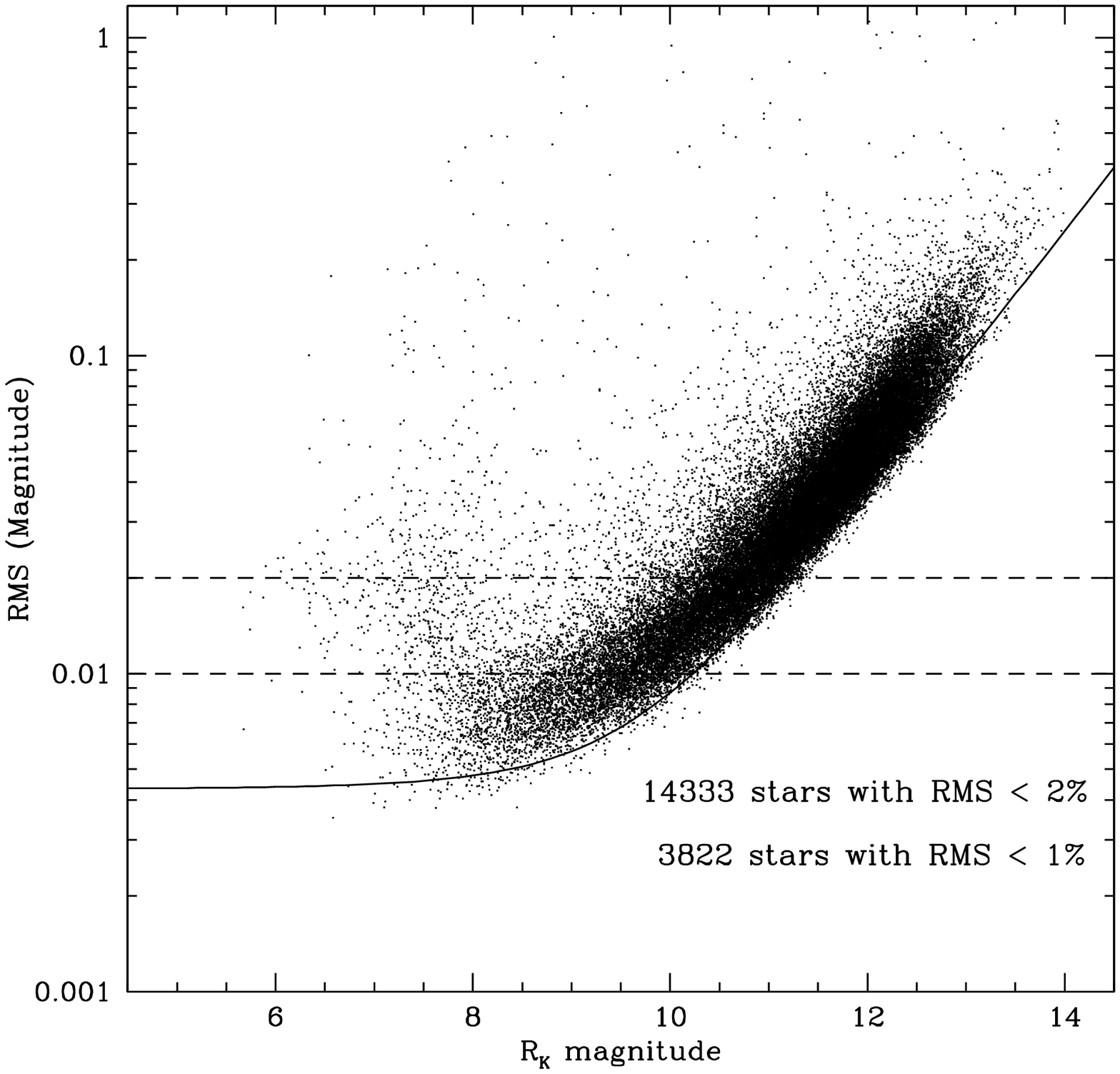}
\caption{RMS plot for eight nights of data from the 80\,mm lens.  Data are 
shown for 49,376 stars.  The dashed lines show the limits for 1\% and 2\% RMS.  The 
solid line represents a noise model including photon noise and sky noise, along with 
an RMS floor of 0.004 magnitudes.}
\label{fig:srms}
\end{figure}

We perform the same analysis for one of the regular survey fields observed 
with the 80\,mm lens using 239 observations over 8 nights with 150-second 
exposures.  We obtain photometry on 49,376 stars in the range $R_K = 6-14$ mag, and 
plot the data in Figure\,\,\ref{fig:srms}.  We find 14,333 stars with
RMS\,$<0.02$\,mag, and 3,822 stars with RMS\,$<0.01$\,mag, with systematics dominating 
for stars brighter than $R_K = 7.5$\,mag.

Overall, the RMS performance is mostly as expected from Poisson statistics 
except at the bright end.  The best precision just reaches the theoretical 
noise limit, with a spread of RMS values above that limit due to 
real-world effects.  For the brightest stars in our data we see a floor in 
which the RMS no longer decreases as the stars get brighter, and instead becomes 
roughly constant at RMS = 0.004 magnitudes.  The RMS floor is indicative of 
a fixed pattern noise component, caused by 
intrapixel sensitivity on the CCD.  The Kodak KAF-E series CCDs
are 2-phase front-side illuminated devices in which the second poly-gate
electrode on each pixel is a transparent gate made of Indium-Tin-Oxide
(ITO) to boost the overall quantum efficiency of the device
\citep{meisenzahl2k}.  Over much of the wavelength regime of interest for
KELT, the ITO material is $\sim$2 times more transparent than the
regular silicon oxide material used on the first poly-gate.  The result
is significant pixel substructure in which the quantum efficiency varies
stepwise across each 9\,$\mu$m pixel, which introduces a component of fixed-pattern
noise that produces the observed RMS floor.  We note that more recent models of
commercial CCD cameras with the Kodak KAF-16801 detector are using a
newer version of this device that incorporate a front-surface microlens
array that mitigates the intrapixel step in transmission, but persons
contemplating similar systems to our own should be aware of the issue
and take it into account.  

We do not expect to obtain high-precision photometry for the very 
brightest stars in our data, but the RMS floor is well below the 1\% level and 
it should not significantly affect our ability to detect transits.  We plot 
noise models in Figures\,\,\ref{fig:crms} and \,\,\ref{fig:srms}, which include 
photon noise and sky noise, along with the RMS floor.  In the future we will 
choose a lens focus that makes slightly larger FWHM images to minimize these effects. 

\section{Representative Results} \label{sec:lcurves}

\begin{figure}
\plotone{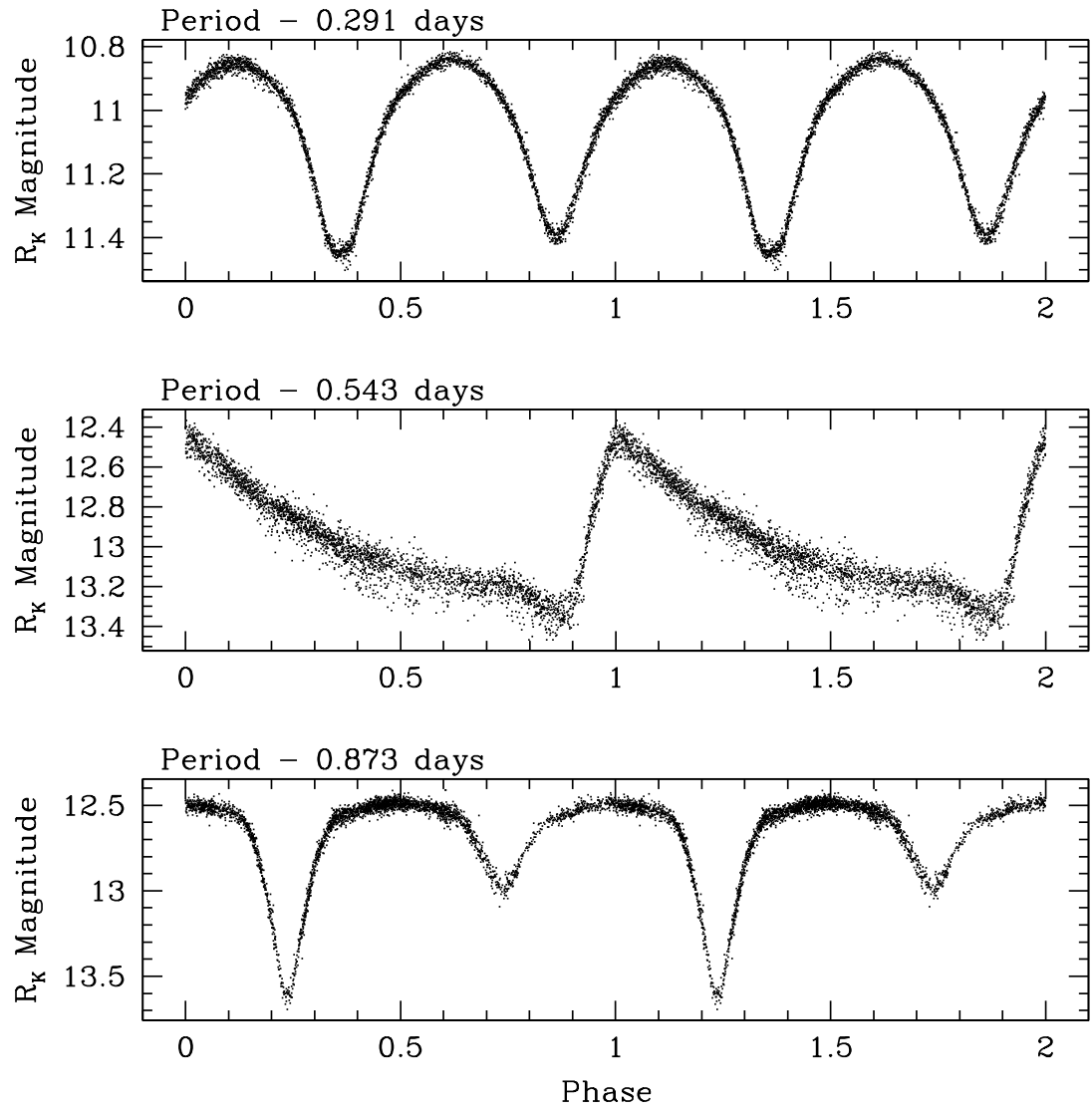}
\caption{Three variable stars discovered with the 200\,mm lens in the 
field of the Praesepe open cluster.}
\label{fig:vars}
\end{figure}

The RMS plots shown in Figures\,\,\ref{fig:crms} and\,\,\ref{fig:srms} 
demonstrate that our telescope can obtain precision relative 
photometry, with large numbers of stars measured at the 1\% level.  
However, simply looking at the RMS information does not 
prove that the data set can yield the consistent quality with low 
systematics necessary for a transit search.  To illustrate that we can fulfill 
that requirement, we show in Figure\,\,\ref{fig:vars} the lightcurves of 
three sample variable stars we have discovered using the 200\,mm lens 
while observing the field of Praesepe.  Even with the night-to-night variations discussed 
in \S \ref{sec:imqual}, we are able to achieve the consistent data quality 
that transit detection requires, and are able to clearly see features 
in the phased lightcurves of a few percent or less.

An even better example of our ability to detect transits can be seen 
in Figure\,\,\ref{fig:dips}.  Here we show two objects we detected in our 
data from the Praesepe field which exhibit transit-like dips in their 
lightcurves.  These effects can be clearly seen at the level of a few 
percent.  These objects are not planetary: follow-up spectroscopy indicates that 
the top object is an F star with a transiting M dwarf companion, and the bottom 
object is a grazing eclipsing binary \citep{latham07}.  However, they 
demonstrate that we can confidently detect 
transit-like behavior at the 1\%-2\% level with our telescope.  A full catalog 
of the variable stars and transit candidates from the KELT observations of the
Praesepe field will appear in a forthcoming paper (J. Pepper, in preparation).

\section{Summary and Discussion}

The KELT project has been acquiring data since October 2004.  We observed the 
field of the Praesepe open cluster with the 200\,mm lens for two months in 
early 2005, and have spent the rest of the time using the 80\,mm lens for a 
survey of 13 fields around the Northern sky.

The KELT telescope, used in ``Survey Mode'' with the 80\,mm lens, reflects the 
design specifications called for in the theoretical paper \citet{pep03}.  The 
performance of the telescope as described in this paper provides a real-world 
evaluation of the potential for this telescope to detect transiting planets.  In 
addition to the lightcurves and RMS plots that show the telescope's abilities, we 
find that the total number of photons acquired from a fiducial $V=10$ mag star 
over an entire observing run is well above the number assumed in \citet{pep03} 
with the parameter $gamma_0$.

This paper has described the KELT instrumentation and performance, with
both the 200\,mm lens used for observing clusters and the 80\,mm lens
used for conducting the all-sky survey.  It is the widest-field
instrument that is currently being employed to search for transiting
planets, and we observe brighter stars than other wide field surveys.  The 
performance metrics demonstrate that it is capable of
detecting signals at the $\sim$1\% level needed to detect Jupiter size planets 
transiting solar-type stars.  Further refinements of our reduction process 
promise to expand our sensitivity to transits, such as applying detrending 
algorithms of the sort developed by \citet{tamuz05}.  We also plan to conduct 
a full analysis of red noise (i.e. temporally correlated systematic 
noise, see \citet{pont06} for a full description) for all KELT data.

\begin{figure}
\plotone{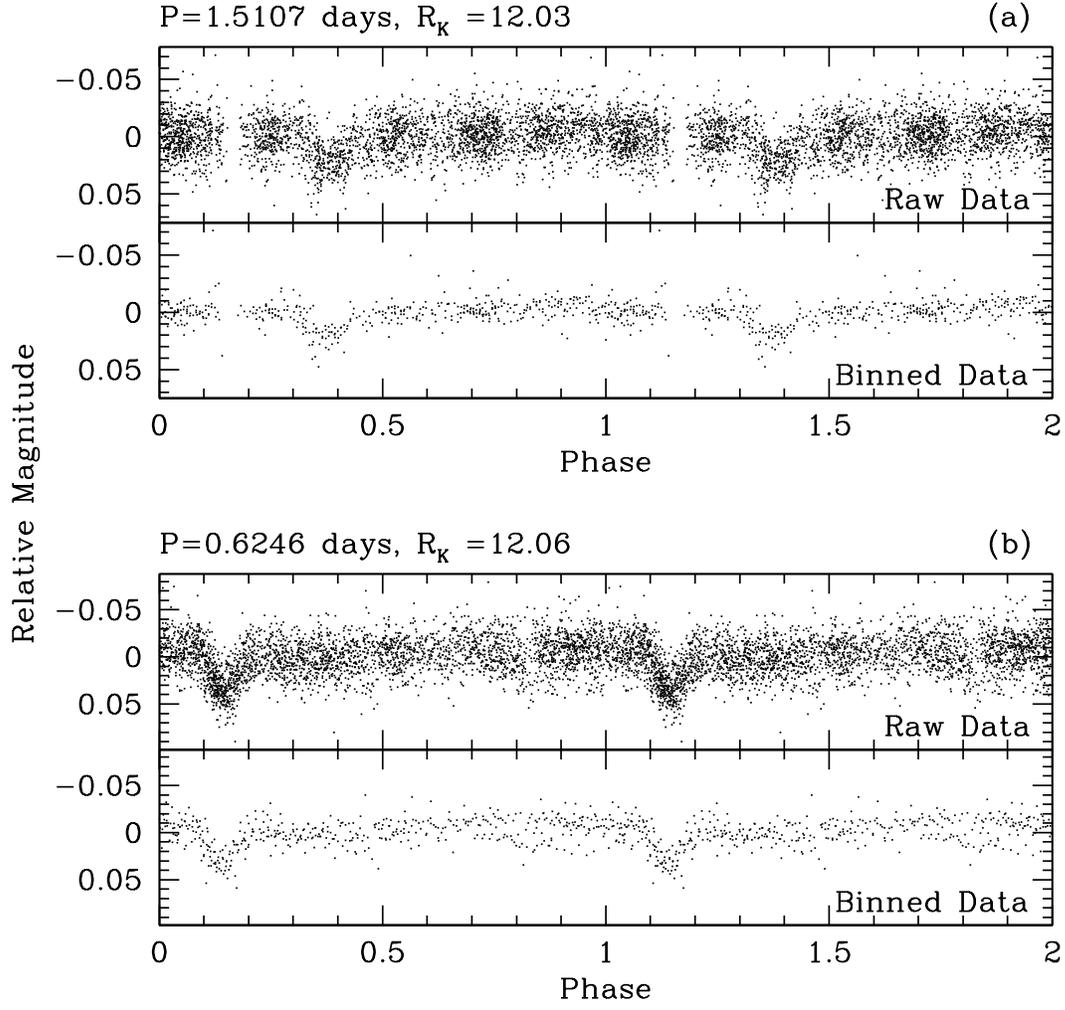}
\caption{Two transit candidates discovered with the 200\,mm lens in the field of 
the Praesepe open cluster.  The lower panel of each plot shows the data binned 
in 10-minute bins.  Follow-up spectroscopy indicates that object (a) is an F star 
with a transiting K dwarf companion, and object (b) is a grazing eclipsing binary.}
\label{fig:dips}
\end{figure}

To date, we have detected over 100 previously unknown variable stars in our 
observations towards Praesepe, and we have identified several lightcurves with 
transit-like behavior, of which we show two in Figure\,\,\ref{fig:dips}.  Future 
papers will report on the success in discovering variable stars and searching 
for planets in Praesepe, along with the full transit search for the all-sky survey.

\acknowledgments 

We would like to thank the many people who have helped with this research, 
including Scott Gaudi, Andrew Gould, Christopher Burke, and 
Jerry Mason.  We would also like to 
thank Apogee Instruments and Software Bisque for supplying the camera and 
mount for the telescope and for excellent service for the hardware.  This 
work was supported by the National Aeronautics and Space Administration 
under Grant No. NNG04GO70G issued through the Origins of Solar Systems program.

\clearpage

\clearpage

\clearpage

\clearpage

\clearpage

\clearpage

\clearpage

\clearpage

\clearpage

\clearpage

\clearpage

\clearpage

\clearpage

\end{document}